# Tuneable molecular doping of corrugated graphene

D. W. Boukhvalov

*Computational Materials Science Center, National Institute for Materials Science,*

*1-2-1 Sengen, Tsukuba, Ibaraki 305-0047, Japan*

*Density functional theory (DFT) modeling of the physisorption of four different types of molecules (toluene, bromine dimmer, water and nitrogen dioxide) over and above graphene ripples has been performed. For all types of molecules changes of charge transfer and binding energies in respect to flat graphene is found. The changes in electronic structure of corrugated graphene and turn of π-orbitals of carbon atoms in combination with chemical structure of adsorbed molecules are proposed as the causes of difference with the perfect graphene case and variety of adsorption properties of different types of the molecules. Results of calculation suggest that the tops of the ripples are more attractive for large molecules and valley between ripples for small molecules. Stability of molecules on the ripples and energy barriers for migration over flat and corrugated graphene is also discussed.*

E-mail: D.Bukhvalov@science.ru.nl

# 1. Introduction

Graphene is two dimensional carbon allotrope [1, 2] with unusual physical [3] and chemical [4, 5] properties. Functionalization is the route for tune the various properties of graphene. Recently discussed limits for the covalent functionalization of graphene [5-7] makes noncovalent graphene functionalization important for the further applications. Sensitivity of graphene for the adsorption of single molecule was experimentally observed [8]. Further theoretical studies of physisorpition of different molecules at graphene considered changes of graphene electronic structure [9-12] possible solubility of graphene [13] or noncovalent functionalization of graphene by different biomolecules [14].

Our previous works [4, 7] suggest for the enormous enhancement of chemical activity of corrugated graphene [15-17] caused by the changes in atomic and electronic structures of the carbon scaffold. Difference between adsorption properties of flat graphene and single wall carbon nanotubes [18] suggest importance of taking into account curvature of the carbon substrate for correct description of physisorpition process. In mentioned works about physisorpition of various molecules on graphene sheet only several possible stable positions were explored. Experimentally observed migration of carbon chain over graphene [19] requires paying attention not only for the most stable position of adsorbent but also for migration barriers. Ripples [20, 21] and charge impurities [22, 23] are discussed as source of carrier densities inhomogeneities. It requires exploration of the interactions of ripples with sources of charges.

In present work dependence of binding energy and charge transfer between adsorbed molecules and graphene are explored. For these studies four different molecules have been chosen. First is the water which is usually adsorbed on graphene at ambient conditions and also penetrate between graphene and $SiO_2$ substrate [11]. Second molecule is $NO_2$ which is experimentally [8] and theoretically [9] studied as graphene dopant. Third type of molecule is bromine dimmer which is using for production of graphite intercalated compounds [24] and proposed for graphene production [25]. Last of studied molecules is toluene ($C_6H_5CH_3$). This molecule used for solution of carbon based compounds and have the chemical composition closer to the solvents proposed for graphite solution and delamination [26, 27]. For bromine and toluene molecules migration barriers were calculated for examine the role of atomic structure to fluidity of adsorbent on graphene flat.

## 2. Computational method and model

The modeling is carried out by the density functional theory realized in the pseudopotential code SIESTA [28], as was done in our previous works [4-7]. All calculations are done with using of the local density approximation (LDA) [29]. An applicability of used for description of layered systems is discussed in Ref. [30] (and refs. therein). All calculations were carried out for energy mesh cut off 360 Ry and k-point mesh 4×4×2 in Mokhorst-Park scheme [31]. Graphene with ripples is not strictly two dimensional system [7]. The ratio of the width of used supercell to the height of the ripples with shemisorbed specie is about 4:1. Thus the using more than one k-point in z direction is required. During the optimization, the electronic ground state was found self-consistently using norm-conserving pseudo-potentials for cores and a double-$\zeta$ plus polarization basis of localized orbitals for carbon and oxygen, and double-$\zeta$ basis for hydrogen. Optimization of the bond lengths and total energies was performed with an accuracy of 0.04 eV/Å and 1 meV, respectively.

For the theoretical modelling of ripples graphene supercell 8×8 (128 carbon atoms in unit cell) have been chosen (Fig. 1). This supercell is identical to used for studying of chemisorption on ripples [7]. For build the model of the ripple all atoms in the centre of the supercell have been smoothly shifted and several atoms on the top is fixed. Several atoms on the edges of supercell are also fixed on initial (zero) height of graphene flat. Further optimisation of atomic structure provides formation of smooth surface of rippled graphene similar to obtained within molecular dynamic simulations [32]. For the calculation of physisorpition on the ripples were used optimized atomic structure of the ripples with given height with the same immovable atoms of carbon scaffold. The modelling of physisorpition on the different sides of ripples (see Fig. 2 a-c) was also performed.

The binding energy defined as $E_{bind} = E_{ripple+molecule} - (E_{ripple} + E_{molecule})$, where $E_{ripple}$ is the energy of initial ripple of the same height and $E_{molecule}$ is the energy of single molecule of adsorbent in the empty box. All used energies were calculated for optimized atomic structures. The charge transfer defined as difference between total number of electrons on valent orbital of studied molecule and calculated occupancies of same orbitals after adsorption. The migration energy defined as difference between the total energies at initial and current steps of migration (see for example Fig. 2d).

## 3. Migration of the molecules over flat and corrugated graphene

Previous studies of migration of weakly bonded (without formation of covalent bond) species over graphene had explored only single adatoms migrations [33]. For this case the calculation only two possible intermediate positions are required for estimate migration barrier. The study of the migration of molecules contained several atoms require another scheme of calculation including several intermediate steps. For calculate the total energy of graphene with physisorbed molecule at current step of migration two atoms in molecule was fixed above several points on graphene sheet (see Fig. 3). The distance between molecule and carbon substrate is the same as for totally optimized initial structure. The initial step of migration corresponds with the most stable configuration of physisorbed molecule on graphene (for $Br_2$ molecule – both of the bromine atoms over the centres of carbon hexagons). The final step of migration process is appropriate the same position of molecule over graphene shifted to the one migration step (4.26 Å for bromine and 8.52 Å for toluene). Results of calculations for $Br_2$ and toluene molecules present on Fig. 4. For both molecules the migration energies over flat graphene are smaller that binding energies (245 meV for toluene and 40 meV for bromine [25]). These values are in qualitative agreement with experimental results for high fluidity of bromine molecules over graphene flat [25] and slow migration of carbon chain on graphene scaffold [17]. For the move of toluene molecule methyl tail (-$CH_3$, see Fig. 2a and d) make the energy curve asymmetric (Fig. 4a) in contrast to symmetric curve of bromine molecule migration (Fig. 4b). Structural anisotropy of molecules provides significant limitation for the possible migration pathways in contrast with single adatoms.

The ascent of molecules at the top of the ripple requires much higher energies. For both types of pysisorbed species were found that final (Fig. 2a) and intermediate (Fig. 2d) is stable. The energy required for the removal from this position on the top and 360 meV of the ripple is relative higher. Molecules were moved to the top of the ripples due to Brownian motion caused by annealing will be staying there at the ambient conditions. The similarity of migration curves of bromine and toluene molecules suggest for the bigger role of the shape of the ripple than molecule composition in contrast to migration over flat graphene.

## 4. Charge transfer and binding energies for the molecules above and under the ripples

Stability of molecules on the top the ripples caused by the migration barriers require study of the charge transfer dependency from ripple height and chemical composition of molecule. Results of the calculations for the ripples of various shape presented on Fig. 5. For toluene, bromine and water molecules has been found rise of the charge transfer in respect to flat graphene. For $NO_2$ molecule the situation is opposite. The cause of described phenomena is in changes in electronic structure of corrugated graphene [7, 21, 34] which provides subtle redistribution of charge near the top. Diminishing of the charge per carbon atom is about 0.005 electrons observed in the area with radius 5Å on the top of the ripple. Described electron deficiency near the top enhances the value of charge transferred from donor molecule to graphene. For acceptor molecules such as $NO_2$ the values of charge transfer decrease.

In contrast to adsorption of toluene at convex graphene ripple (Fig. 2a) for concave ripple of same height (Fig. 2b) the charge transfer decay are calculated. For understanding this phenomena and diversity of binding energy curves for different types of used molecules taking into account changes in atomic structure of corrugated graphene is required. In toluene molecule on flat graphene distance between carbon atoms in *meta* position is about 2.50 Å (in respect to 2.46 Å in graphene). This subtle mismatch between the lattice parameters of graphene and interatomic distances in molecules play significant role in the physisorpition process. For the convex ripples π-orbitals of carbon atoms take place in the side (as it was in details discussed for the carbon nanotubes [35, 36]) and better overlap with π-orbitals of carbon atoms in toluene is obtained. For the concave shape of the ripple the distance between π-orbitals in graphene decrease and charge transfer from toluene also diminished. In contrast with relatively big molecule of toluene for nitrogen dioxide both convex and concave shapes of the ripples provide reduce of charge transfer due to decay of electronic density near the top. But for the binding energy between $NO_2$ decreases of distance between π-orbitals of concave graphene provide better overlap and grown of binding energy. In contrast with strong acceptor $NO_2$ the water is weak donor and discussed above effect provide enhancement of charge transfer for the case of adsorption on the ripples with concave shape without changes in binding energy.

Subtle charge redistribution in corrugated graphene provides the changes of charge transfer between graphene and adsorbed molecule. The usual for graphenic systems presence of water molecules (see Ref. [10] and reference therein) also provides charge redistribution in the system. For examine the role of this combination calculation of coadsorption of toluene and water molecule from both sides of ripple (Fig. 2d) has been performed. The distance between molecules and graphene and sum of binding energies stay the same as in case of single molecule adsorption but the charge transfer significantly increase (for toluene from 0.018 to 0.050 and for water molecule from 0.004 to 0.016 electrons, this values correspond to 7.2, 20.0, 1.6 and $6.4 \cdot 10^5$ electrons per $cm^2$ respectively). The joint presence of water and other adsorbent or presence only water on graphene samples could be the key for the understanding several sample depending results for graphene.

**5. Conclusions**

Density functional theory modelling of adsorption properties for different molecules on corrugated graphene suggest for the significant changes in charge transfer and binding energies in respect to flat graphene. Two main causes of variety of both parameters are subtle charge redistribution in corrugated graphene and light turn of π-orbitals on the top part of the ripple. Combination of this causes make ripples of convex shape more attractive for large molecules and ripples of concave shape for the small molecules (2-3 atoms). Light decay of charge near the top of the ripples makes it attractive for the donor molecules, small acceptor molecules more probable distribute in the valley between ripples. Obtained results permit to build on graphene stable at ambient conditions due to examined huge migration barriers areas with *n*- and *p*-types of doping.

**Acknowledgements**

I am grateful to A. K. Savchenko (University of Exeter) for fruitful discussions.

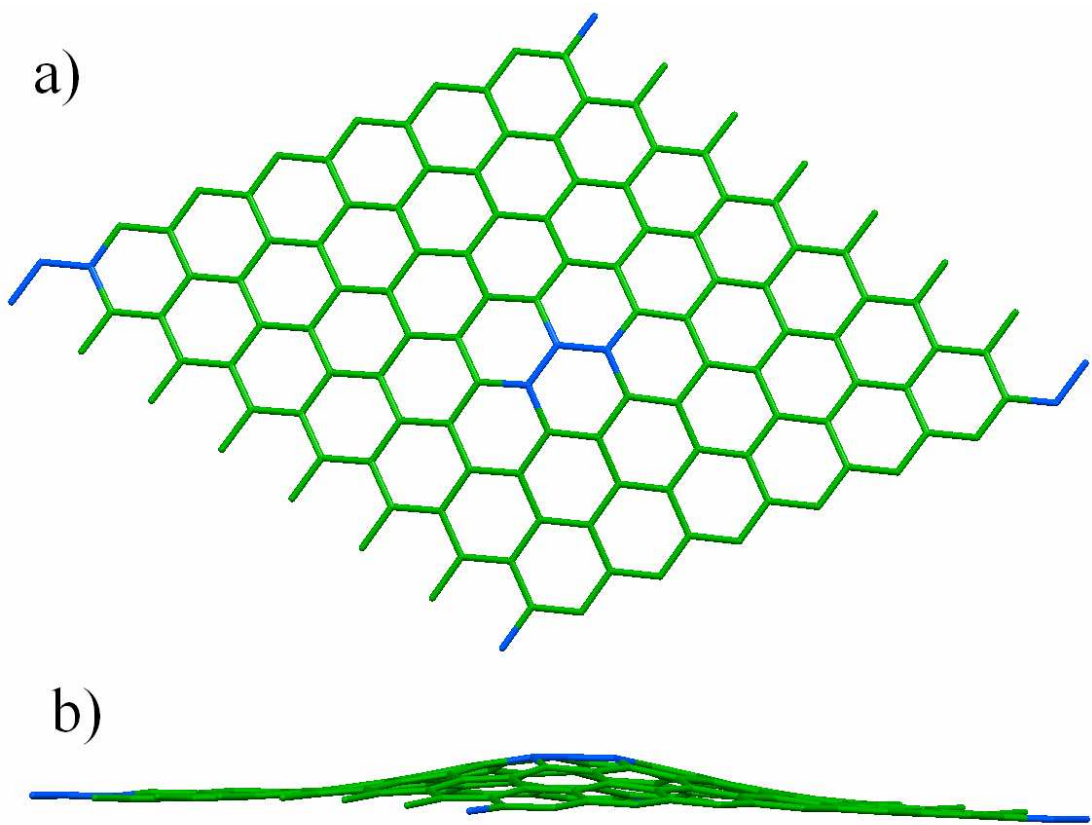

**Figure 1.** Top (a) and side (b) view of the optimized atomic structure of graphene ripple with height 1.478 Å. The fixed carbon atoms are shown by blue colour.

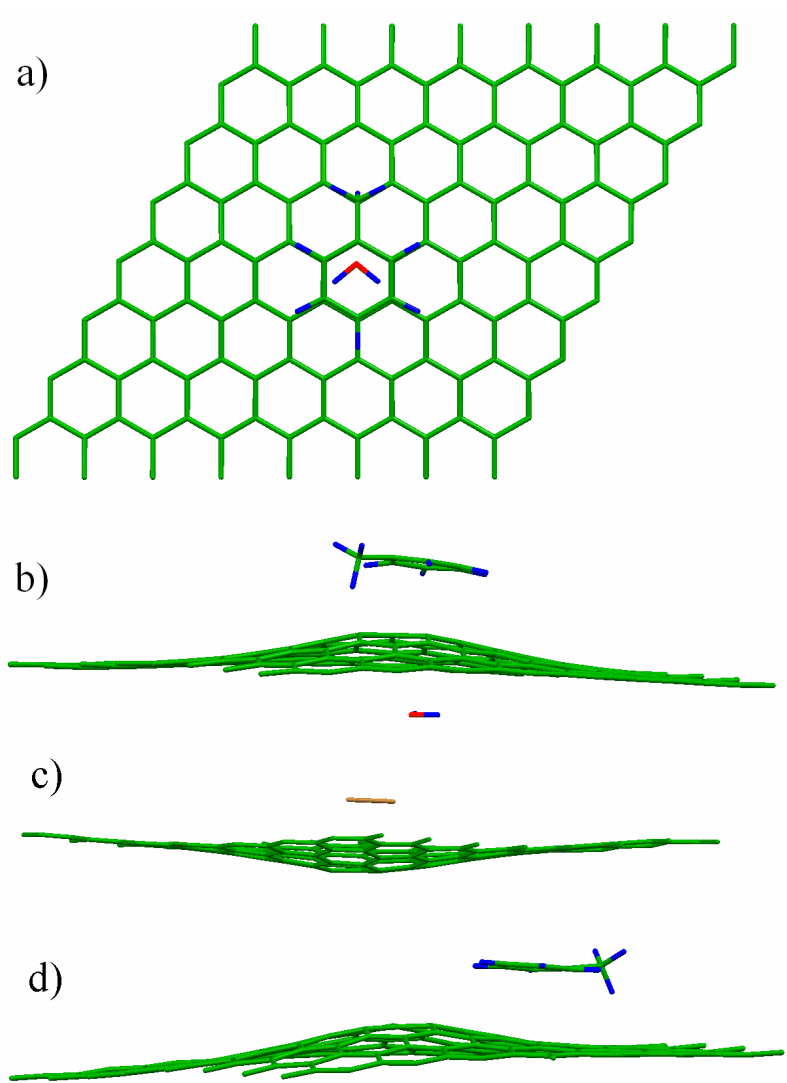

**Figure 2.** Top (a) and side (b) view of optimized atomic structure of graphene ripple with height 1.478 Å with physisorbed toluene molecule from top of the ripple and water molecule from another side. Optimized atomic structures of the same ripple with physisorbed over the valley bromine molecule (c); and toluene molecule in the intermediate step of the movement to the top of this ripple (d). Carbon atoms are shown by green, oxygen by red, bromine by brown, and hydrogen by blue.

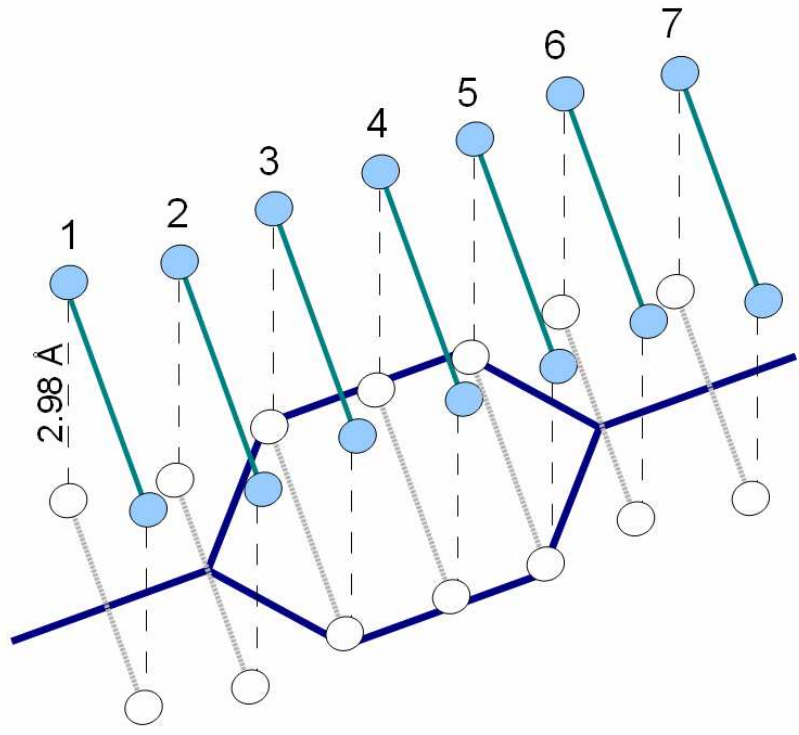

**Figure 3.** A Sketch of the steps of bromine molecule migration over graphene sheet.

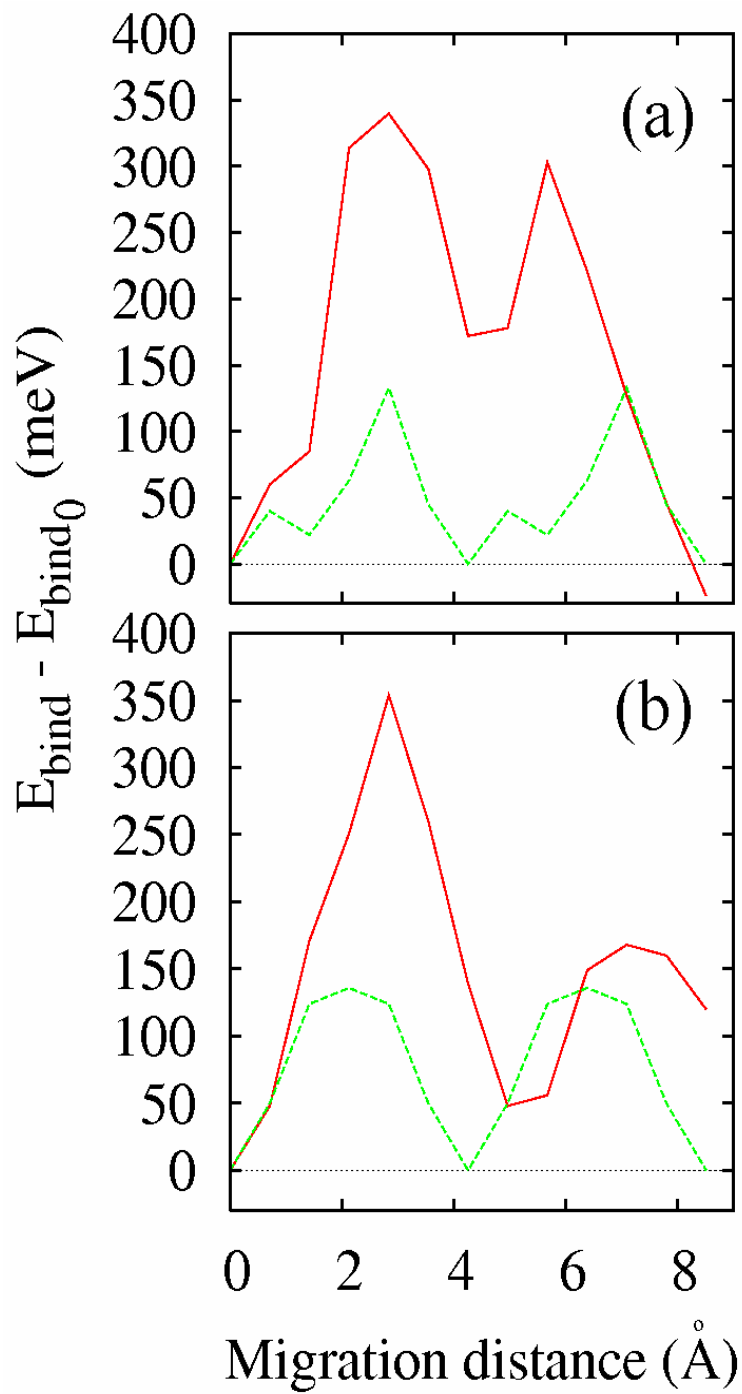

**Figure 4.** Difference between total energies at current and initial steps of migration of toluene (a) and bromine (b) molecules over rippled (solid red line, see also Fig. 2d) and flat graphene (dashed green line).

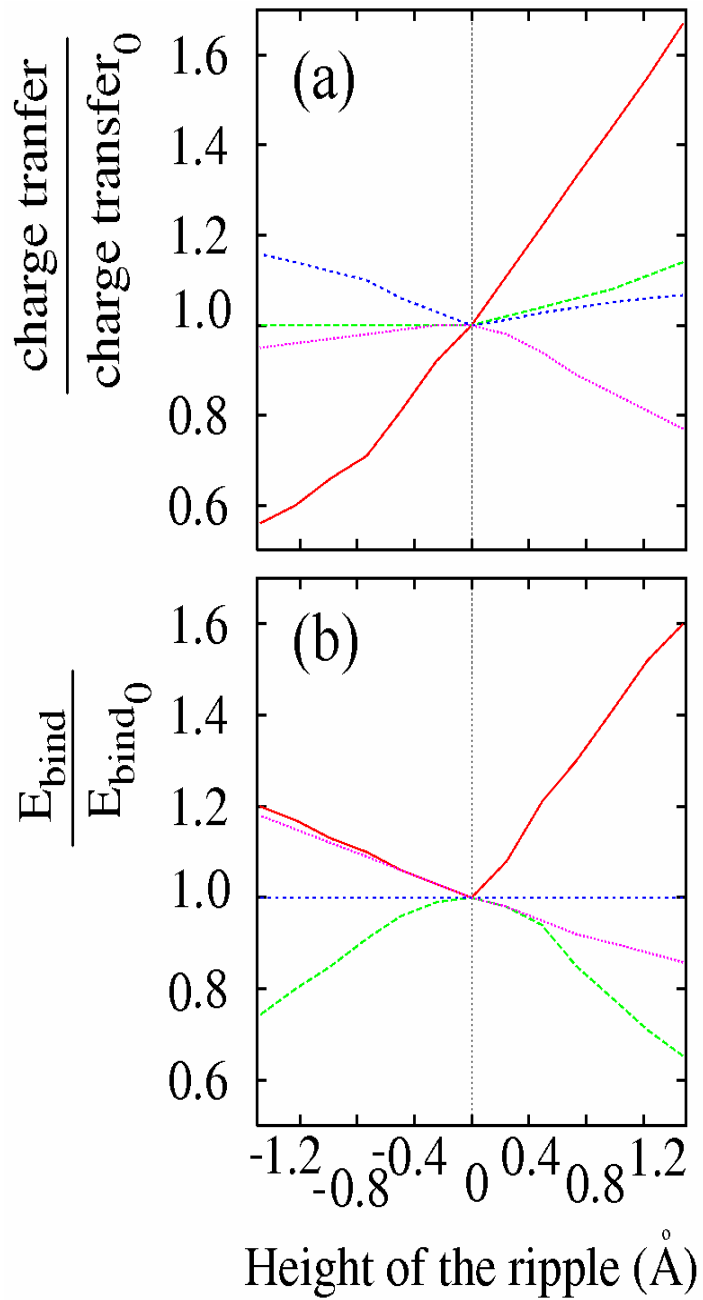

**Figure 5.** Ratio of charge transfer for the case of the ripple with given shape to the case of flat graphene (a), ratio of binding energy for the case of the ripple with given shape to the case of flat graphene (b) for the toluene (solid red line), bromine (dashed green line), water (dotted blue line), and nitrogen dioxide (dotted violet line).